# Infodemiological Study Using Google Trends on Coronavirus Epidemic in Wuhan, China

Artur Strzelecki (✉), Mariia Rizun
University of Economics in Katowice, Katowice, Poland
artur.strzelecki@ue.katowice.pl

**Abstract—**The recent emergence of a new coronavirus (COVID-19) has gained a high cover in public media and worldwide news. The virus has caused a viral pneumonia in tens of thousands of people in Wuhan, a central city of China. This short paper gives a brief introduction on how the demand for information on this new epidemic is reported through Google Trends. The reported period is 31 December 2020 to 20 March 2020. The authors draw conclusions on current infodemiological data on COVID-19 using three main search keywords: coronavirus, SARS and MERS. Two approaches are set. First is the worldwide perspective, second – the Chinese one, which reveals that in China this disease in the first days was more often referred to SARS then to general coronaviruses, whereas worldwide, since the beginning, it is more often referred to coronaviruses.

## 1 Introduction

As at 10 am Central European Time on 20 March, 2020, 245,484 cases globally had been reported of pneumonia caused by the novel coronavirus now known as COVID-19 [1], of which 10,031 had been lethal [2]. The most infected countries are China (n=81,250), Italy (n=41,035) and Iran (n=18,407) [3], [4]. The number of confirmed cases worldwide has exceeded 200, 000. It took over three months to reach the first 100, 00 confirmed cases, and only 12 days to reach the next 100, 000 [5]. There have been 86,035 reported recovered cases in160 countries and territories.

The first set of measures was taken in Wuhan (China), where at 10 am Beijing time on 22 January, 2020, the international airport was closed. Presently, all the countries affected by COVID-19 have closed their borders partially or completely [6].

The coronavirus, currently named COVID-19, belongs to a large family of viruses that can cause many different infections, from a cold to acute respiratory failure syndrome [7]. The one from Wuhan is only the 7th that is known to harm people [8]. There is no single effective way to fight coronaviruses, but only measures to help relieve their symptoms. The symptoms include fever, respiratory problems and lung infiltration [9].

Infodemiology is the area of scientific research that focuses on scanning Internet, publicly available data and other sources for user-contributed health related content [10]. In recent years a lot of research has been done using data collected from Google Trends (GT), Google Flu Trends or Google Health API. Recently there has been a growing number of research using GT [11]. Before it was released, early studies were done on Google Flu Trends – a source for keywords connected to diseases [12].

Google Trends is the source of reversed engineered data. It shows what was searched in Google; this data is normalized in terms of search frequency and presented in relative search volumes. Data is segmented into years, months, days, and into geographical regions. Researchers are able to compare maximum 5 keywords using segments in one try. Studies on GT can be divided into four areas: infectious diseases, mental health, other diseases and general population behavior [13]; and are mainly used to examine the seasonality [14].

The motivation behind this short paper is to use Google Trends data to analyze how search demand for information about coronaviruses is shared worldwide. There is no doubt that people all over the world express personal interest in this issue by searching information and using search keywords.

## 2 Materials and Methods

Methodology used in this study is based on [15]. Data for this research is collected from Google Trends (https://trends.google.com/trends) and is normalized. The highest interest in search keywords is expressed by 100, whereas lack of interest or insufficient data are expressed by 0. GT contains data from different geographical locations, segmented into countries, territories and cities; it also allows to set custom time range. Queries are collected from five specialized search engines: Web, Image, News, Google Shopping and YouTube Search.

Data for the study was retrieved starting from the first mention of this disease noted by Google, which was 31 December, 2019. Data comes from Web Search with two geographical settings: 1) worldwide, to see the interest in coronaviruses over the world; 2) only China. Since Google does not operate in China freely, the authors see the necessity to have it examined separately.

## 3 Results

The first approach is to collect data on the topics: *Coronavirus*, *SARS* and *MERS*. Figure 1 shows the worldwide interest (in search results) from 31 December, 2019 to 17 March, 2020. The data from Google Trends was retrieved on 20 March, 2020. Since there is usually a gap of two days prior to the day of retrieval, the last available data is from 18 March, 2020. Figure 1 shows that information outbreak of interest in *coronavirus,* as well as in *SARS,* started on 20 January, 2020.

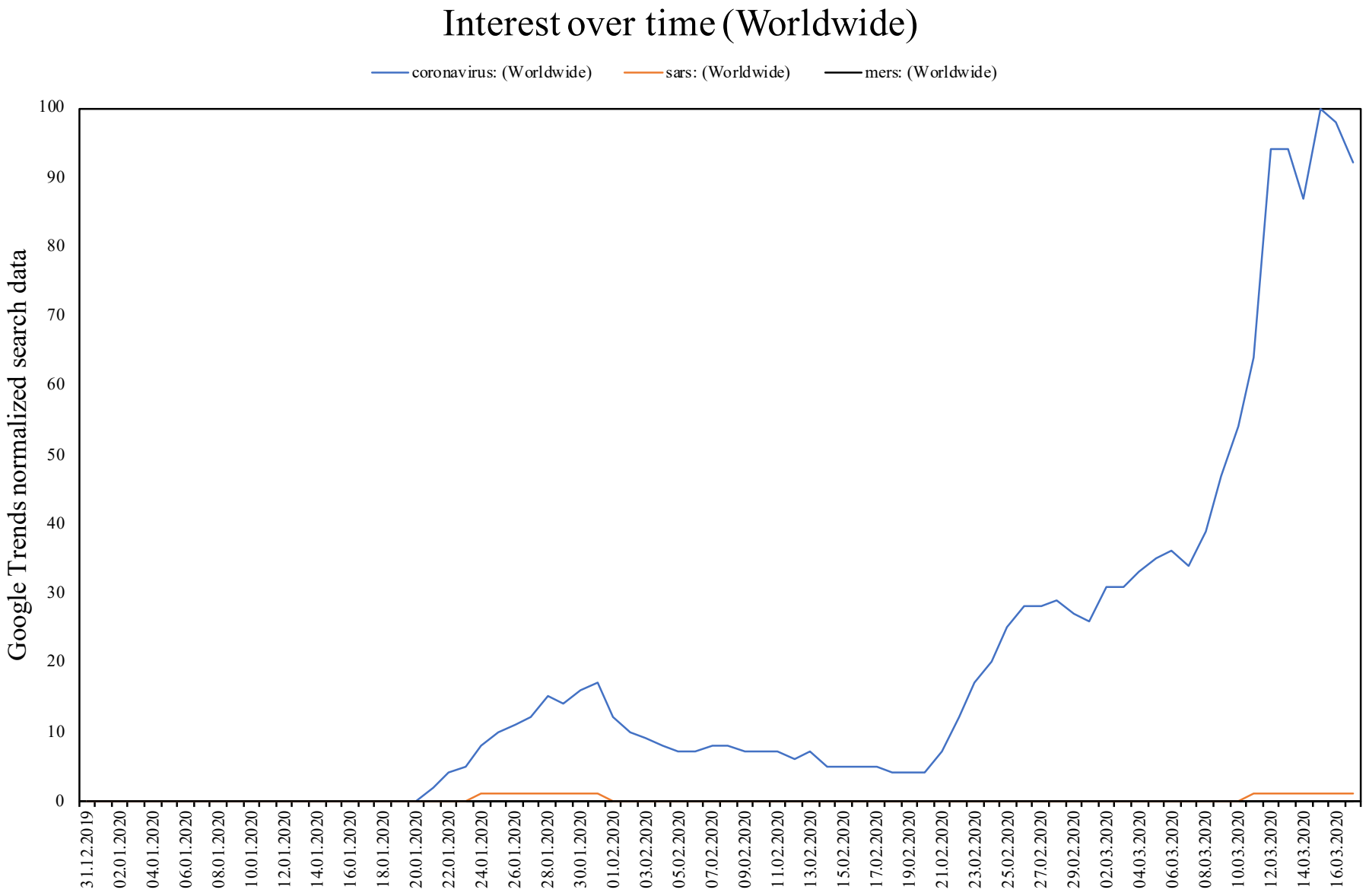

**Fig. 1.** Worldwide interest over time for *coronavirus*, *SARS* and *MERS*; source: Google Trends

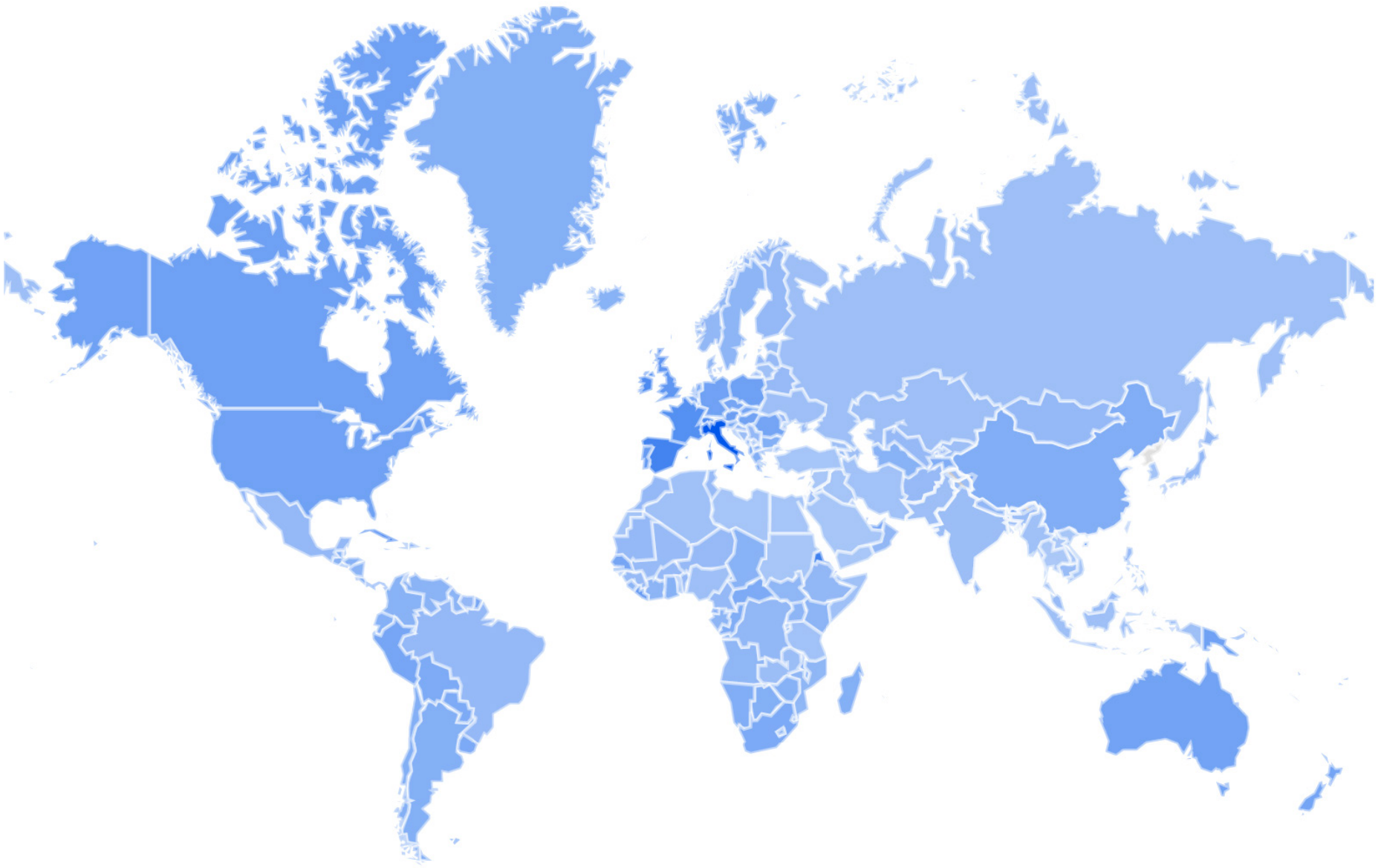

**Fig. 2.** Compared breakdown by countries in period from 31 December, 2019 to 17 March, 2020 for the search term: *coronavirus*; source: Google Trends

While Figure 1 represents how interest changed with time, Figure 2 illustrates how the interest in *coronavirus* varied in different parts of the world. Saturation of blue color in the figure defines the level of interest, with the more intense color meaning higher interest. There is no surprise that the highest interest in the virus was expressed by internet users in Italy, Spain and France. As of 17 March, 2020, interest in *coronavirus*, *SARS* and *MERS* was expressed in 232 countries, according to Google Trends web search data. Search term popularity is relative to the total number of Google searches performed at a specific time, in a specific location.

The second approach of the study, as stated before, is to collect similar data, but only for China. Because of the general unavailability of Google in the China, the data could be less precise. Figure 3 reveals that Chinese search queries on this disease were concentrated around *SARS* keyword in the beginning. The trend changed on 22 January, when interest in *coronavirus* had become greater than that in *SARS*. Referring to Figure 2, we can state that the overall trend in search queries is the same for China and the other countries – after 22 January *coronavirus* became the topic of highest concern all over the world. At the same time, in China, as well as globally, the end of January is characterized by decrease of interest in the issue. It can be assumed that at that time the world was more or less aware of the problem, so no new queries were generated by internet users.

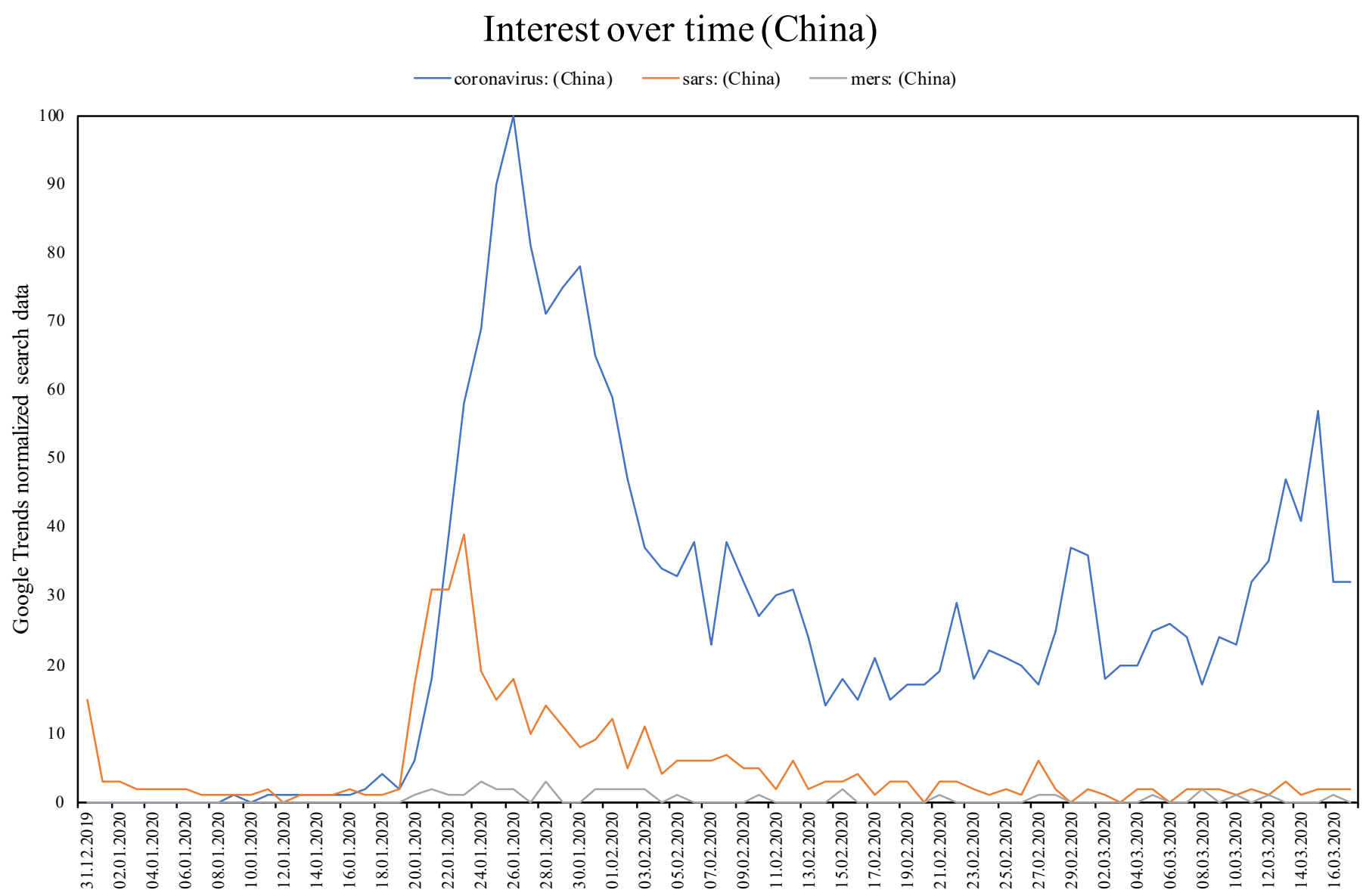

**Fig. 3.** Interest over time for *coronavirus, SARS and MERS* in China; source: Google Trends

Figure 4 focuses on the interest in SARS among the population of China. The search data was retrieved for 28 Chinese subregions – the ones where people were more threatened by the virus. Saturation of blue color in the figure shows the highest interest expressed in Wuhan (the center of virus outbreak), where the color is the most intense, and in the provinces around it. However, it is necessary to remember that Google is officially unavailable in China and this results in limited data from Google Trends in China.

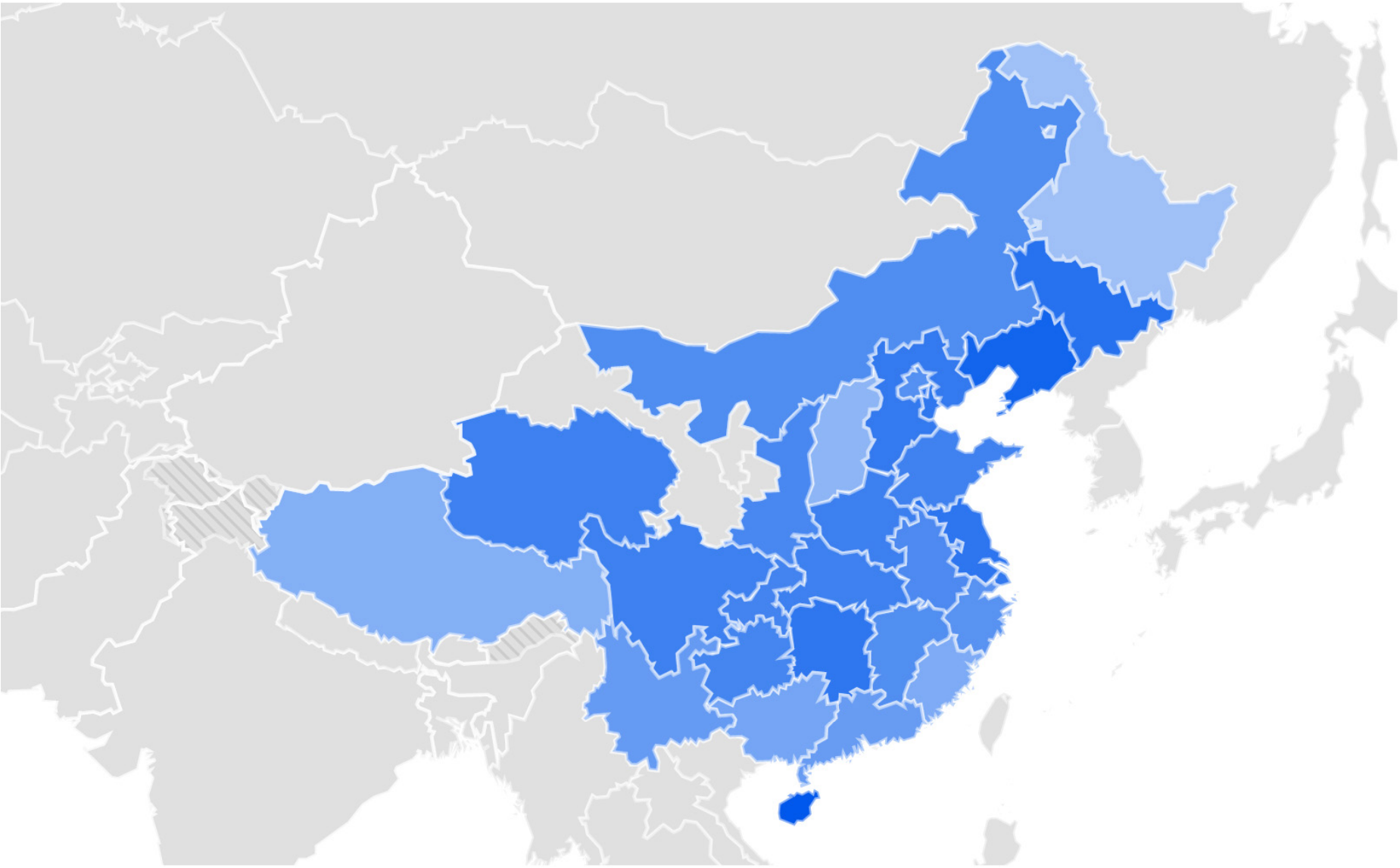

**Fig. 4.** Compared breakdown by Chinese regions in period from 31 December, 2019 to 18 March, 2020 for the search term: *SARS*; source: Google Trends

Figure 5 presents the Chinese and global search results compared to the number of new COVID-19 cases. Since GT has a data delay, GT data at the time of writing ends on 14 March. The left axis shows normalized GT search volume. The right axis shows new COVID-19 cases [4]. The data interval is one day.

The situation worldwide has changed since a rapid increase in cases was reported in South Korea, Italy and Iran. GT data reveals the rapid growth of the second wave of interest in coronavirus since 21 February, 2020. This rising interest trend is observed worldwide and in the countries, where a rapid increase in cases of laboratory-confirmed COVID-19 has been reported since 21 February 2020 [11]. It is visible on the trend line that the increasing interest is slightly slowing between 13 and 15 of March and then increasing again. Data reveals that currently the highest peak of interest in coronavirus was on 15 March, 2020.

It is clear that the interest in coronavirus changes together with aggravation of the situation around the world. However, Figure 5 also shows that the interest among Chinese users started to decrease regardless the new cases that were still appearing. The situation worldwide, to the reported date, is different. Yet, considering the example of China, we can assume that the growth of interest might slow down as well as the overall awareness around the world grows.

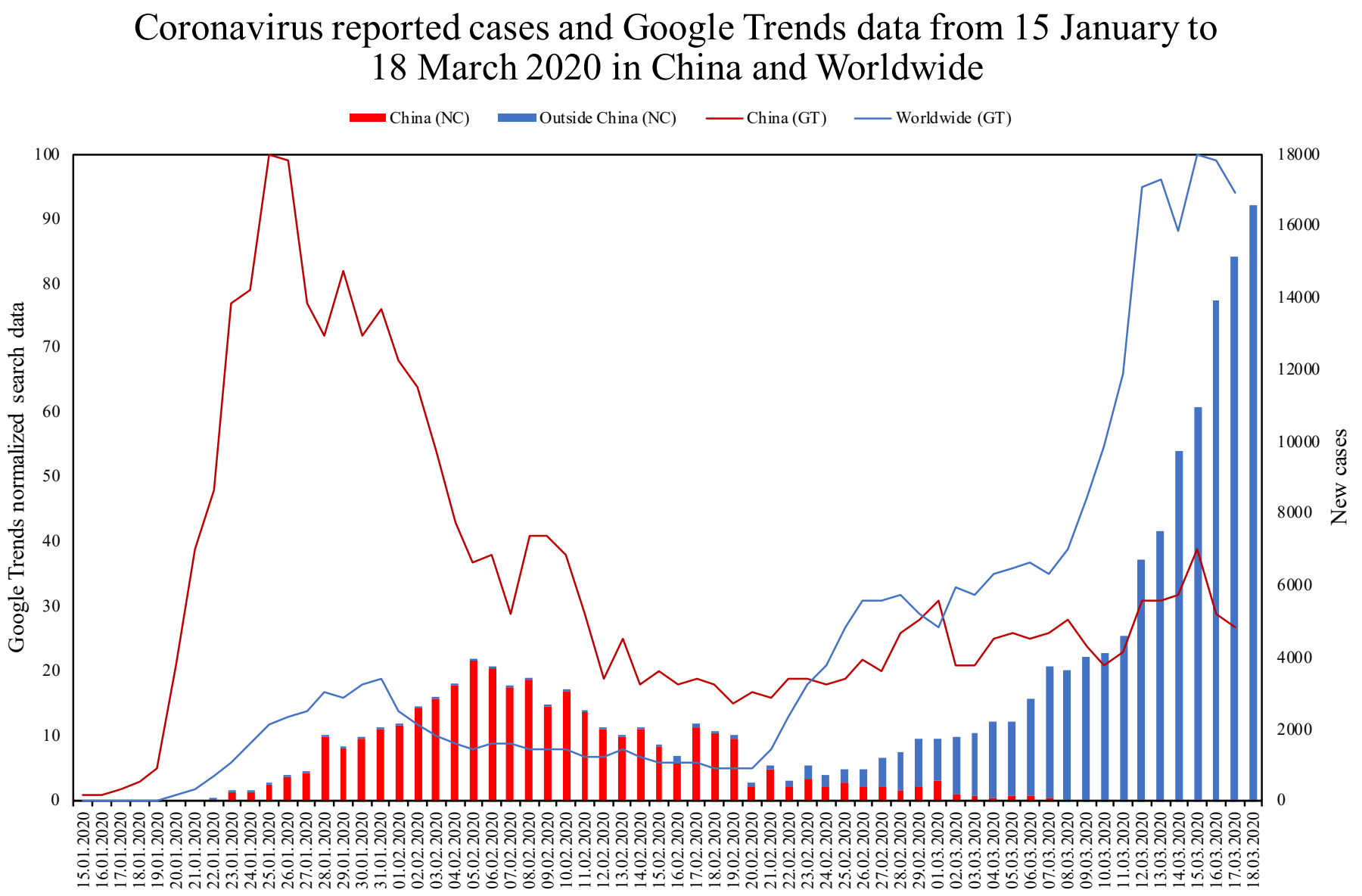

**Fig. 5.** Coronavirus reported cases and Google Trends data from 15 January to 18 March, 2020 in China and Worldwide.

## 4 Discussion

China has a strong collective memory of the SARS outbreak in 2003 [16], which caused an eventual 8,098 cases, resulting in 774 deaths reported in 17 countries, with the majority of cases in China mainland and Hong Kong [17]. The epidemiological similarity between this year's outbreak and that of SARS is striking. SARS was then traced to animal markets [18] similarly like in case of COVID-19. This could explain why in China greater interest is observed in *SARS* than in the general term *coronavirus*. However, after 22 January, 2020, volume of interest in coronavirus surpassed that in SARS. It is certainly connected with the rapid release to the public of the genome sequence of the new virus by Chinese virologists [19].

In Figures 1 and 3 the values are calculated on a scale from 0 to 100, where 100 is the day with the most popularity as a fraction of total searches in that day, the value of 50 indicates the day which is half as popular. The value of 0 indicates the day when there was not enough data for this term. Table 1 presents data only from 10 Chinese regions. However, it is important to stress that because of general unavailability of Google in China, people there use Baidu for search; this results in low precision of data from Google Trends in China.

Data from Google Trends reveals interest from many countries worldwide. Actually, interest is shown in at least 232 countries, whereas outbreak is currently reported in 177 countries, except China. The numbers confirm that currently this situation is of high concern globally. This short paper is a starting point for further analysis of information demand spread across search engines. Currently Google offers Trends service, which acts as a reverse data engineering and allows to collect data on peoples' interest, which, in this case, is the interest in coronavirus (COVID-19) epidemic.

The key finding of this research is that Google Trends forecasted the rise of new cases. In the first wave, new cases were increasing day-by-day for 6 days after the highest peak of GT worldwide interest. In the second wave, interest in coronavirus on GT is still rising, which predicts the increasing number of new cases reported daily. This implies that national health services should implement additional health measures against countries other than China.

The limitation of the study is the lack of Google Trends data about China because of the general unavailability of Google in China. Another limitation is that data about the countries presently affected by the virus is changing rapidly every day, thus results are only relevant to the reported date.

## 6 Authors

**Artur Strzelecki** is with University of Economics in Katowice, Katowice, Poland (artur.strzelecki@ue.katowice.pl).

**Mariia Rizun** is with University of Economics in Katowice, Katowice, Poland (mariia.rizun@ue.katowice.pl).